# On the Role of Service Concept in IT

**Ioan Despi**
**University of New England, Armidale, NSW, Australia**
**Lucian Luca**
**„Tibiscus" University, Timisoara, Romania**

**ABSTRACT.** Hard times affecting world-wide economy have strong consequences and are challenging IT departments from all sorts of enterprises. Expensive software projects are replaced by component-based agile systems and paradigms like SOA, REST, cloud computing are the new buzz-words. Behind the canvas, the service concept plays a central role, which we try to reveal.

## 1 Introduction

We are in the middle of the largest labor force migration in the history, as society moves from manufacturing goods to producing services. In 2005, 70% of US GDP came from service business, while in countries like China or India it represented 35% [RP05]. Through collaborative partnerships enterprises are now global virtual organizations which deliver goods and services at a high competitive price. To stay here, they have to reassess their operational and organizational structures to meet new requirements, such as innovation, flexibility and shorter deadlines. As a consequence of this globalization, manufacturers might quickly lose their competitiveness and therefore adapt by learning that the only hope is their service components which would distinguish themselves from other players in the same competition.

Services sectors today cover everything, from traditional ones (transport, logistics, distribution, etc.) to contemporary ones (e-commerce, financial engineering, consulting, etc.). The concept is evolving and new high-value areas are recognized as services on a daily basis (post-sales training, IT outsourcing, on-demand consulting, etc.). In this scenario, IT plays a critical role, as the principal facilitator for geographically dispersed manufacturing and services delivery. From an end user's point of view, IT is just a service and as the economy moves towards an information services based one, a better understanding of the service concept on all its aspects (e.g., marketing, innovation, design, engineering operations, management, etc.) would be beneficial.

New exciting areas of research include service marketing, design and engineering, modeling and innovation, operations and management. In a service-oriented enterprise, IT is a service based on service oriented architecture (SOA), component process model (CPM), business process management (BPM) and web services (WS).

## 2  The Service Concept

The polysemantic word 'service' is too broad and vague. The service is not originally a technical concept as it is borrowed from the field of business. There, a service provider is offering to do something that will benefit sorme others called service consumers. The providers offer to contract with consumers these things, so the consumers know in advance what they will get for their money.

In the marketing field, a service is described as "an activity or benefit that one party can offer to another that is essentially intangible and does not result in the ownership of everything" [KA05]. Some authors also mention that a service is perishable, non-transportable and heterogeneous [Lov92], while others mention the soft separation between services and products [V+85]. Sometimes a service is provided as a collection of related services ('service flower') or as an aggregation of simpler services ('bundling') [BAG03].

The service quality is estimated by comparing performance versus expectations. A high quality service is one for which performance is according to expectations [Pa+84]. A customer of a service should know in advance what quality of service she can expect. For this to happen, the expectations are to be defined, usually by means of Service Level Agreements (SLA) negotiated between the provider and the customers of a service. This can prevent later conflicts between them, as both parties agreed on the properties and performance of a service [Thi05]. A SLA is realized accordingly to the service management lemniscate [TBZ04] which actually shows the gap between the needs of the consumer and what a supplier is prepared to deliver.

Some authors [TBZ04] distinguish between a service pit and a service shell. The pit is formed by the core (IT) object delivered to the customer, while the shell contains the supporting processes and agreements of a service.

## 3  The Main Technologies

Nowadays business environments change faster and faster so businesses are required to be agile if they want to respond quickly to changing market needs and opportunities. The IT architectures should be flexible and agile to support change and actually, they have continuously evolved and in the past twenty years or so. A common way for IT to keep pace with change was via client/server and n-tier architecture styles which implemented a component-based development (CBD) approach, where software that supports businesses was built from discrete entities of functionality that can be reused or reconfigured in order to adapt to changing environments.

Examples of such technologies include CORBA, Enterprise JavaBeans (Sun), COM+, .NET (Microsoft). All of them are based on defining reusable objects with a strictly specified interface, such that if somebody wanted to use a component, she needed to know in advance what components were available and what interfaces they could export.  When more and more components became available, it started to be harder to find an appropriate one. Because components have been produced by different vendors, usually they were incompatible.

SOA and web/cloud services replace today large scale enterprise systems as well as CBD. They represent a distributed solution that is event-driven, flexible and cost saving. They are service-based architectures built on modular components, reusable functionality and interoperability which are deployed because they allow enterprises to respond faster to business changing requirements. The idea is to pack processes as a network of interoperable and repeatable self-contained services to achieve modularity, flexibility and agility such that new capabilities can be built into the current system as needed. The functionality is dynamically located, invoked based on specific criteria, and communicated in messages. Components are intensively reused so the costs are down and efficiency increases.

Based on XML standards (such as WSDL, SOAP, UDDI), web services (WS) comprise a family of related standards working together to deliver interoperability between different platforms and languages by means of passing messages from one service to another and not to bother on the underlying architecture of applications. In other words, WS refer to the technologies that allow for making connections between different services, and so a service can be seen as a endpoint of a connection, while a standard is an approach for defining, publishing and using web services. Specifications for web services are described by Web Services Description Language (WSDL), Web Services Policy Framework (WS-Policy), Web Services Dynamic Discovery (WS-Discovery), Web Services Metadata Exchange (WS-MetadaExchange), Web Services Endpoint Language (WSEL). The common way to develop state of the art services to implement SOA is to use Sun products, such as Java WebServices Developer Pack, Java 2 Platform Enterprise Edition, Servlets and Enterprise Java Beans.

Cloud services use the concept of interoperable services plus a virtualisation component to relieve internal servers from being overloaded by the constant reuse of these services within the system. The approach is based on using Internet to store and process information permanently in the cloud and caching it only temporarily on client machines.

The SOA concept has many different interpretations in literature but there is a consensus that SOA is not an architecture by iteslf but it is more a philosophy and an architectural template, providing opportunities to create service oriented architectures [L+07]. Some authors consider SOA as a meta-concept, used to describe an architecture of resources [Gar03], which can be services or organisational resources. Because SOA uses services to integrate Business with IT, it is more than a technological architecture and more than an overview of business processes.

SOA vision is based on the design paradigm of CBD, saying that functionality can be split over components that are loosely coupled. The architecture is usually represented as a three-layers pyramid [O+03]. At the bottom layer basic components (services) offer discrete parts of functionality. SOA solves the problem of scalability by providing a mechanism for the publication and discovery of these services, in contrast with CBD. The basic operations are discovery, service publication, interface description and message exchange. The aggregation of multiple basic services into composite services is done at the composition layer. This second layer also offers support for coordination, monitoring of performance and quality of service. The top layer is the management layer, with Service Level Agreements (SLA), assurance and support for the information systems architectures.

One enterprise embraces SOA thinking that eventually it will build and enable a distributed computing infrastructure across the internet, such that organizations can collaborate and integrate their applications. For this promise, they voluntarily offer all of their applications to the public by means of a centralized registry, which is usually shared between a number of companies operating in the same industry sector. Discovery of services is a crucial aspect of SOA and the availability of services will be dynamic such that everybody can change his mind quickly between different suppliers of a specific service. The service can be located in all parts of the enterprise, in contrast with Client/Server or OOPD approaches, where they belonged only to IT department.

The main benefit of enabling SOA is obvious: using services provided by others reduces overall development and maintenance costs. Then, it offers opportunities for incremental development, deployment, maintenance and extension of business applications by reusing business components in other business processes [Gar03]. More, SOA can be implemented in parallel to existing architectures and systems and can use these by encapsulating parts of them as services. The limitations of this technology have been highlighted in [L+07].

A policy is a design rule together with its enforcement. A crucial factor for the success of a SOA is governance, the development and enforcement of SOA policies. SOA governance refers to activities and procedures related to exercising control over services in an SOA environment. It features everything a company uses to ensure SOA is done in accordance with best practices, architectural principles, government regulations and pertinent laws [Me08]. SOA governance includes a large palette of aspects and to make it easier, one has to use a service register to centralize the service specifications and governance policies. SOA governance is as much about organizational issues and how people work together to achieve business goals as it is about any technology [All06, H+06]. SOA is first and foremost about the design of the business, not the technology; it is as much a state of mind as it is a technology; it is as much about behaviour and orientation as it is about programming per se [HF07].

## 4  The Service Concept in SOA

In the past years, service-orientation has become a pivotal design approach with commonly accepted principles. The meaning of the concept service orientation depends on the employee's role in the company [Kin05]. A technician sees this as a flexible option to create dynamic, cooperating and loosely coupled applications. The business people and managers consider service orientation as a model for collecting information and business logic from various systems into a single interface. The business analyst can describe and map empolyees, external capabilities providers and automation systems into a single model by using service orientation concept [Kin05]. A service in the context of SOA is defined in different ways by different authors [Erl05, KA05]. The most important features are:

A service is a black box about an observable behaviour delivering a certain function that is well-defined, self-contained, and implementation independent.

It performs a specific task, typically a business one (e.g., validating a bank card) by means of a well-defined interface (what) described in an implementation (how) independent manner which shows the API to the service together with all the details needed for a client to invoke the service. More, it does not depend on the context or state of other services.

A service is loosely coupled, as it is coarse-grained and uses asynchronous message exchange, and it can be dynamically discovered. Because the client and the service operate (usually) in different environments, a service can be invoked remotely. The coupling between services is done through a directory, which is a registry of available interfaces. The directory is searchable both by users and componenets.

The environment of the service has a constraint role, expecting a certain behaviour from the service. In principle, services are not allowed to update other systems. The degree in which the service accomplishes these expectations partially determines the quality of the service.

Service identification is a combination of techniques (like domain decomposition, existing asset analysis, goal-service modeling) originated from separation of concerns that can be used top-down, bottom-up or mixed. For example, at the business level the top-down approach is most appropriate, so one can start to identify and specify business services by using business use cases and business process modeling.

A service in the context of SOA is not necessarily a concept in the context of marketing (an intangible benefit offered by one party to another [KA05]) but has similarities if seen from a service provider's perspective: the model has to define who offers the service (the provider), what the service performs (a functional description), and on which channels the service is available [B+05].

Service management is a completely organizational approach that states the value of the service (to the organization) and the utility of service (from the user's point of view) [Al88]. The most common way to manage services is the concept of applications services portofolio [B+05], where the services are categorized in four basic areas (strategic, key, operational, support, and high potential) depending on their current of future contribution to the success of the business.

Service quality is normally evaluated based in accordance to requirements and expectations. Whether or not the service meets the participant's (customer, employee, stakeholder, etc.) expectations is established by two tests of quality, technical quality and emotional quality [B+05].

Service reuse is the main promise and reason why SOA is so popular in enterprises in the last years, as it could bring cuts in costs and faster deliverables. As everybody wants to use a service to help accomplish another service, services must encapsulate factored functionality from the design phase. Because services are pieces of business processes, they naturally involve people, information, and systems. As a consequence, when defining a service, one is structuring and organizing business processes and business organizations as well as systems. Thus, if services are to be reused, they must fit and align with different business processes and hence they require the total architecture perspective and active business involvement [Bro07]. Theoretically, all services within a SOA can become universally reused, saving on development and maintenance costs and boosting productivity. Practically, reuse is only partial with a small degree of regularity and only at lower organizational layers [Ben06, H+06].

## 5  Conclusions

When enterprises invest money in services they do this to best fulfill both long-term and short-term goals. Often services are abstractions of business processes, therefore hard to describe and evaluate costs and benefits. Because of the complexity of services and difficulties in understanding them, enterprises do not treat services as any other investment. A better understanding of the nature of services and their features will lead to better IT services and accomplishment of enterprise's goals.